# Simulation of non-ideal gases and liquid-gas phase transitions by lattice Boltzmann equation

Xiaowen Shan*

*Center for Nonlinear Studies, Los Alamos National Laboratory, Los Alamos, NM 87545*

Hudong Chen†

*Department of Physics and Astronomy, Dartmouth College, Hanover, NH 03755-3528*

## Abstract

We describe in detail a recently proposed lattice-Boltzmann model for simulating flows with multiple phases and components. In particular, the focus is on the modeling of one-component fluid systems which obey non-ideal gas equations of state and can undergo a liquid-gas type phase transition. The model is shown to be momentum-conserving. From the microscopic mechanical stability condition, the densities in bulk liquid and gas phases are obtained as functions of a temperature-like parameter. Comparisons with the thermodynamic theory of phase transition show that the LBE model can be made to correspond exactly to an isothermal process. The density profile in the liquid-gas interface is also obtained as function of the temperature-like parameter and is shown to be isotropic. The surface tension, which can be changed independently, is calculated. The analytical conclusions are verified

---

*Also at Department of Physics, Dartmouth College, Hanover, NH 03755-3528

†Present address: Exa Corporation, 125 Cambridge Park Street, Cambridge, MA 02140

by numerical simulations.

## I. INTRODUCTION

Since the Lattice Gas Automaton (LGA) was introduced [1–3] as an alternative method for simulating fluid flows, a great amount of effort has been devoted to developing models for various physical systems [4]. In a LGA model, both space and time are discrete and particles move along links on a regular lattice and collide with each other on each lattice site according to some properly designed rules and conservation laws. A set of fluid equations can be derived describing the macro-dynamics of the system which, under proper limitations, lead to incompressible Navier-Stokes equations.

Although the LGA offers a simple and efficient tool for simulating some fluid flows, it is difficult to achieve high accuracy with the LGA because of its statistical noise. Moreover, it usually suffers from other problems including a lack of Galilean invariance and velocity-dependent pressure, unless some special procedures are taken to eliminate them [5]. To overcome these problems, an alternative approach, known as the lattice Boltzmann Equation (LBE) method [6–8], was derived from the LGA. In this method, rather than following dynamics of the particles, a Boltzmann equation governing the evolution of the particle distribution function is directly solved. The solution, after some simple mapping, can be shown to obey the desired fluid equations under certain conditions [9,10]. While having inherited the major advantages of the LGA in simulating fluid flows, such as the computational simplicity and the ease in handling complex boundary conditions, the LBE method is free of the problems mentioned above.

There have been LGA models proposed for simulating flows with multiple thermodynamic phases. One of the first models was constructed by Chen *et al.* [11] in which the particles interact according to a nearest-neighbor potential. The thermodynamic properties are equivalent to that of the Ising model so that phase transition and the critical point is exactly known [12]. However, the momentum exchange in this model does not correspond to that in a real fluid. It can not be used for simulating realistic fluid flows.

Alternatively, Appert *et al.* [13–15] proposed a momentum conserving LGA model for simulating the liquid-gas phase transition. In their model, particles over several lattice sites can interact with each other. As the range of the interaction exceeds a certain critical value, a liquid-gas type of phase transition is observed. Nevertheless, this model suffers from the shortcomings of the usual LGA models, especially the lack of Galilean invariance which becomes severe when there is a large density variation. Moreover, the surface tension in this model depends on the orientation of the interface with respect to the lattice structure [15].

In a previous paper [16], we presented the basic idea of a new LBE model for modeling flows with multiple thermodynamic phases based on nearest-neighbor interactions. In addition to the advantages of the LBE models, it offers some features highly desirable in simulating multi-phase flows. First of all, it is computationally efficient since for most purposes, the non-locality can be restricted to nearest-neighbor interactions. The additional computation involved in such a non-local interaction is equivalent to that of a streaming step, which can be efficiently implemented on parallel computers. Secondly, it can handle fluid systems composed of arbitrary number of components, each of which can have different mass and transport coefficients. Third, the equation of state is exactly expressed in terms of the inter-particle potential, and can be tuned to precisely match any given functional form. When the equation of state is properly chosen, a liquid-gas phase transition can occur. Critical phenomena can be studied since the critical point is analytically calculable.

In this paper, we focus our attention to the application of this model to one-component systems and discuss in detail the equilibrium properties when a liquid-gas phase transition is allowed. In Section II, we present the details of the inter-particle interactions and show the momentum conservation in a global sense. The liquid-gas phase transition is shown to occur when the inter-particle potential is properly chosen and a temperature-like parameter is below a critical value. In Section III, the two-phase coexistence curve is calculated analytically from the microscopic mechanical balance condition. The pressure tensor is given in analytical form, from which the density profile across the liquid-gas interface is obtained. The surface tension is also calculated as a function of the temperature-like parameter. Comparison with thermodynamic predictions reveals that this LBE model corresponds to an isothermal PVT system when the inter-particle potential is chosen to be in a particular form. In Section IV, the analytical results are verified by numerical computation. The agreement between the theory and the simulation is found to be excellent. The density profile is also shown to be isotropic with respect to the underlying lattice structure. Finally in Section V, we offer further discussion and suggest some future research directions.

## II. LATTICE-BOLTZMANN MODEL WITH PAIRWISE INTER-PARTICLE POTENTIALS

We recall the one-component lattice Boltzmann equation with the BGK [9,17] collision term:

$$n_a(\mathbf{x}+\hat{\mathbf{e}}_a, t+1) - n_a(\mathbf{x}, t) = -\frac{1}{\tau}\left[n_a(\mathbf{x}, t) - n_a^{eq}(\mathbf{x}, t)\right], \tag{1}$$

where the distribution function $n_a(\mathbf{x}, t)$ at lattice site $\mathbf{x}$ denotes the particle population moving in the direction $\hat{\mathbf{e}}_a$; $n_a^{eq}(\mathbf{x}, t)$ is a prescribed equilibrium distribution function; and $\tau$ is the collision time [9]. It has been shown [9,10] that if the equilibrium distribution function is chosen to be

$$\begin{aligned} n_a^{eq}(\mathbf{x}) &= n(\mathbf{x})\left[\tfrac{1-d_0}{b} + \tfrac{D}{c^2 b}\hat{\mathbf{e}}_a\cdot\mathbf{u} + \tfrac{D(D+2)}{2c^4 b}\hat{\mathbf{e}}_a\hat{\mathbf{e}}_a : \mathbf{u}\mathbf{u} - \tfrac{D}{2bc^2}\mathbf{u}^2\right], \quad a = 1, \cdots, b; \\ n_0^{eq}(\mathbf{x}) &= n(\mathbf{x})\left[d_0 - \tfrac{1}{c^2}\mathbf{u}^2\right], \end{aligned} \tag{2}$$

at long wave length limit, Navier-Stokes equations with an ideal gas equation of state

$$p = \frac{c^2(1-d_0)}{D}n \tag{3}$$

will be obtained from the kinetic equation (1). The corresponding kinematic shear viscosity, $\nu$, is $c^2(\tau - \frac{1}{2})/(D+2)$ and the bulk viscosity, $\eta$, is $d_0 c^2(\tau - \frac{1}{2})/D$. In the equations above, $n(\mathbf{x}) = \sum_a n_a(\mathbf{x})$ and $\mathbf{u} = \sum_a \hat{\mathbf{e}}_a n_a(\mathbf{x})/n(\mathbf{x})$ are the particle number and the fluid velocity at lattice site $\mathbf{x}$. $b$ is the number of total links connecting a lattice site to its nearest neighbors. The velocity vectors are represented by $\hat{\mathbf{e}}_a$ ($a = 1, \cdots, b$) with magnitudes equal to $c$, the lattice constant divided by the time step. $D$ is the dimension of the chosen lattice [3]. The constant, $d_0$, is the equilibrium fraction of particles with zero speed. It can be easily verified that with the equilibrium distribution function, $n_a^{eq}$, so chosen, the BGK collision operator on the right hand side of Eq. (1) conserves the total particle number, $n(\mathbf{x})$, and the total momentum, $\sum_a \hat{\mathbf{e}}_a n_a(\mathbf{x})$, at each site.

To simulate non-ideal gases and their mixtures, long-range interactions between particles must be included. Since particles then exchange momentum through the long-range attractive or repulsive forces in addition to short range collisions, site-wise momentum conservation must be abandoned. In a previous paper [16], we suggested the inclusion of effects of the inter-particle forces in a system composed of $S$ components by defining a potential of the following form,

$$V(\mathbf{x},\mathbf{x}') = \sum_{\sigma=1}^{S}\sum_{\bar{\sigma}=1}^{S} G_{\sigma\bar{\sigma}}(\mathbf{x},\mathbf{x}')\psi^{\sigma}[n^{\sigma}(\mathbf{x})]\psi^{\bar{\sigma}}[n^{\bar{\sigma}}(\mathbf{x}')], \tag{4}$$

where, $n^{\sigma}$ and $n^{\bar{\sigma}}$ are the number density of components $\sigma$ and $\bar{\sigma}$. Since in a lattice model, particles reside on lattice sites with fixed distance between each other, the number density on each site also determines the average distance between particles. The potential is then made to be proportional to the products of local "effective masses" $\psi^{\sigma}(n^{\sigma})$, which are functions of local densities and are defined individually for different components. Although the exact mapping is currently unknown, the forms of the $\psi$'s control the detailed nature of the interaction potential and determine the equation of state of the system. The forces between different components can be either attractive or repulsive and can have different strengths, as defined by the Green's function, $G_{\sigma\bar{\sigma}}(\mathbf{x},\mathbf{x}')$, which is reduced to $G_{\sigma\bar{\sigma}}(\mathbf{x}-\mathbf{x}')$ in homogeneous systems.

In the case of a one-component system with nearest-neighbor interaction, the Green's function is further reduced to a single number $\mathcal{G}$, namely

$$G(\mathbf{x}-\mathbf{x}') = \begin{cases} 0 & |\mathbf{x}-\mathbf{x}'| > c \\ \mathcal{G} & |\mathbf{x}-\mathbf{x}'| = c\ , \end{cases} \tag{5}$$

which measures the strength of the interactions among the particles on the nearest neighboring sites. The long-range force due to the potential Eq. (4) acting on the particles at site $\mathbf{x}$ is then

$$\mathbf{F}(\mathbf{x},t) = -\mathcal{G}\psi(\mathbf{x},t)\sum_{a=1}^{b}\psi(\mathbf{x}+\hat{\mathbf{e}}_a,t)\hat{\mathbf{e}}_a, \tag{6}$$

where $\psi(\mathbf{x},t) = \psi[n(\mathbf{x},t)]$. To reflect the momentum change caused by this force at each time step, we let $\mathbf{u}$ in the equilibrium distribution function given by Eq. (2) be

$$\mathbf{u} = \frac{1}{n(\mathbf{x})}\left[\sum_{a=1}^{b}\hat{\mathbf{e}}_a n_a(\mathbf{x}) - \tau\mathcal{G}\psi(\mathbf{x},t)\sum_{a=1}^{b}\psi(\mathbf{x}+\hat{\mathbf{e}}_a,t)\hat{\mathbf{e}}_a\right]. \tag{7}$$

The second term is the momentum change due to the long-range force between particles at site $\mathbf{x}$ and its nearest neighbors.

With this choice of interaction, although the momentum is not conserved by the BGK collision operator at each site, it can be easily shown that the total momentum is indeed conserved. At each step, summing over all the lattice sites, we have the following net momentum change:

$$\Delta \mathbf{P} = -\mathcal{G} \sum_{\mathbf{x}} \sum_{a=1}^{b} \psi(\mathbf{x}) \psi(\mathbf{x} + \hat{\mathbf{e}}_a) \hat{\mathbf{e}}_a. \tag{8}$$

By replacing the dummy variable $\hat{\mathbf{e}}_a$ with $-\hat{\mathbf{e}}_a$, the above is also equal to,

$$\Delta \mathbf{P} = \mathcal{G} \sum_{\mathbf{x}} \sum_{a=1}^{b} \psi(\mathbf{x}) \psi(\mathbf{x} - \hat{\mathbf{e}}_a) \hat{\mathbf{e}}_a. \tag{9}$$

If the boundaries do not introduce any momentum flux into the system, then $\mathbf{x}$ can be viewed as a dummy variable. Let $\mathbf{x}' = \mathbf{x} - \hat{\mathbf{e}}_a$ and then drop the primes, the above equation becomes

$$\Delta \mathbf{P} = \mathcal{G} \sum_{\mathbf{x}} \sum_{a=1}^{b} \psi(\mathbf{x} + \hat{\mathbf{e}}_a) \psi(\mathbf{x}) \hat{\mathbf{e}}_a = -\Delta \mathbf{P}, \tag{10}$$

and hence, $\Delta \mathbf{P} = 0$. In another words, no net momentum change is incurred by the interaction between particles introduced above. Same conclusion can be proved for systems with more than one component.

With the new equilibrium distribution function given by Eqs. (2) and (7), following the same Chapman-Enskog procedure [3], we have previously shown [16] that the Navier-Stokes equation with the equation of state

$$p = \frac{c^2}{D} \left[ (1 - d_0) n + \frac{b}{2} \mathcal{G} \psi^2(n) \right], \tag{11}$$

can be obtained. Fluids with a quite general class of equations of state can then be modeled by selecting the function $\psi(n)$. Once the function, $\psi$, is properly chosen, this equation of state will exhibit many essential features of a liquid-gas phase transition. In the limit of weak non-local force, namely $\mathcal{G} \to 0$, Eq. (11) approaches the equation of state of an ideal-gas given by Eq. (3) with a constant temperature proportional to $\frac{c^2}{D}(1 - d_0)$. As the parameter $-(1 - d_0)/\mathcal{G}$ decreases below a critical value, (either by increasing the inter-particle force or increasing the rest particles at each site), pressure $p$ becomes a non-monotonic function of $n$, qualitatively similar to the van der Waals equation of state for many choices of $\psi$, e.g., $\psi(n) = 1 - \exp(-n)$ [16]. The unphysical negative compressibility $dp/dn$ corresponds to a thermodynamic instability and the system will segregate into a denser (liquid) phase and a lighter (gas) phase. The critical value of $(1 - d_0)/\mathcal{G}$ will be given by the following two equations:

$$\frac{\partial p}{\partial n} = \frac{c^2 \mathcal{G}}{D} \left( \frac{1 - d_0}{\mathcal{G}} + b\psi\psi' \right) = 0, \quad \text{and} \quad \frac{\partial^2 p}{\partial n^2} = \frac{bc^2 \mathcal{G}}{D} \left( \psi\psi'' + \psi'^2 \right) = 0. \tag{12}$$

The equation of state (11) gives constant relations between the pressure and the density, similar to the isotherm in a PVT system. The parameter $-(1 - d_0)/\mathcal{G}$ formally plays the same role in this LBE model as the temperature in the van der Waals theory of phase transition. However, it should be noted that the original kinetic equations (1), (2) and (7) do not imply explicit energy conservation. The existence of an energy-like quantity is unknown at this point. As a consequence, this LBE model does not have a well defined statistical mechanical temperature, and in general, the correspondence between the LBE model and an isothermal process cannot be established. A more detailed discussion will be given later in Section III.

## III. EQUILIBRIUM PROPERTIES

It is of great interest to calculate the coexistence curve for a two-phase system. For the PVT system, the coexistence curve is given by a procedure known as the Maxwell construction, which requires that the Gibbs potential is minimized around the phase transition where it is triple-valued. But for the current lattice Boltzmann model, there is not *a priori* an explicit energy conservation relation. Whether one can find an effective free energy for this system is still unknown. However, the simplicity of the form of the inter-particle force in our model enables us to obtain, from the microscopic mechanical balance conditions, the coexistence curve as well as the density profile across the liquid-gas interface. The surface tension can also be calculated once the pressure tensor and the density profile are known.

We consider the one-component lattice Boltzmann model with nearest neighbor interaction as described in last section. The momentum flux tensor comprises a kinetic term due to the free streaming of the particles and an additional potential term from the long-range inter-particle force. For the inter-particle force given by Eq. (6), the momentum flux tensor is

$$\mathbf{\Pi} = \sum_a n_a \hat{\mathbf{e}}_a \hat{\mathbf{e}}_a + \frac{\mathcal{G}}{2}\psi(\mathbf{x}) \sum_a \psi(\mathbf{x} + \hat{\mathbf{e}}_a) \hat{\mathbf{e}}_a \hat{\mathbf{e}}_a, \tag{13}$$

where the first term is the conventional kinetic term and the second is the potential term. We consider the system in equilibrium, namely $n_a(\mathbf{x}, t) = n_a(\mathbf{x})$. We further assume there is no net mass transfer along any of the links connecting two lattice sites. This can be expressed as

$$n_{\bar{a}}(\mathbf{x}) = n_a(\mathbf{x} + \hat{\mathbf{e}}_a), \qquad \text{for} \qquad \hat{\mathbf{e}}_a = -\hat{\mathbf{e}}_{\bar{a}}. \tag{14}$$

For a homogeneous density distribution, the fluid velocities before and after the collision step equal to each other and vanish when the system is in equilibrium. But they are not equal in the interface region due to the momentum change included in the collision step. Neither velocity is zero in this interface zone. However, the magnitude of the fluid velocity of the equilibrium distribution given by Eqs. (2) and (7) can be calculated. Multiply the lattice Boltzmann equation (1) by $\hat{\mathbf{e}}_a$ and sum over $a$. Taking into consider Eq. (14), we have, for steady state, the following equation at each site

$$-n\mathbf{u} - n\mathbf{u} = -\frac{1}{\tau}\left(n\mathbf{u} - n\mathbf{u}^{eq}\right), \tag{15}$$

where $\mathbf{u}^{eq} = \sum_a \hat{\mathbf{e}}_a n_a^{eq}$ is the fluid velocity of the equilibrium distribution and

$$n\mathbf{u}^{eq} = n\mathbf{u} - \mathcal{G}\tau\psi \sum_a \hat{\mathbf{e}}_a \psi(\mathbf{x} + \hat{\mathbf{e}}_a) \simeq n\mathbf{u} - \mathcal{G}\tau \frac{c^2 b}{D} \psi \nabla \psi. \tag{16}$$

We can then solve for $\mathbf{u}^{eq}$ from the equations above and obtain the following equation,

$$\mathbf{u}^{eq} = -\left(\tau - \frac{1}{2}\right) \frac{c^2 b \mathcal{G}}{Dn} \psi \nabla \psi. \tag{17}$$

Substituting the equilibrium distribution given by Eqs. (2) and (17) into Eq. (13), and using the following Taylor expansion of $\psi(\mathbf{x} + \hat{\mathbf{e}}_a)$,

$$\psi(\mathbf{x} + \hat{\mathbf{e}}_a) = \psi(\mathbf{x}) + \nabla\psi \cdot \hat{\mathbf{e}}_a + \frac{1}{2}\nabla\nabla\psi : \hat{\mathbf{e}}_a \hat{\mathbf{e}}_a + \cdots, \tag{18}$$

we obtain the pressure tensor

$$\begin{aligned} \mathbf{p} = &\left[\frac{1 - d_0}{D} c^2 n + \frac{c^2 b \mathcal{G}}{2D} \psi^2 + \frac{c^4 b \mathcal{G}}{4D(D+2)} \psi \nabla^2 \psi\right] \mathbf{I} + \\ &\frac{c^4 b \mathcal{G}}{2D(D+2)} \psi \nabla\nabla\psi + \left(\tau - \frac{1}{2}\right)^2 \frac{c^4 b^2 \mathcal{G}^2}{D^2 n} \psi^2 \nabla\psi \nabla\psi, \end{aligned} \tag{19}$$

where $\mathbf{I}$ is the unit tensor. In a homogeneous fluid where all the derivatives vanish, the pressure tensor is isotropic, and once again we obtain the equation of state Eq. (11).

In order to calculate the density profile across a liquid-gas interface, we consider a flat interface coincident with the $x$-$y$ plane separating the two phases. The densities in the bulk liquid and gas phases are $n_l$ and $n_g$ respectively. We choose the point where $n = (n_l + n_g)/2$ as the origin of the $z$-axis without lose of generality. All the variables vary in the $z$ direction only, so that $n = n(z)$ and $\partial/\partial x = \partial/\partial y = 0$. In this situation, the pressure tensor $\mathbf{p}$ becomes diagonal:

$$\mathbf{p}(\mathbf{x}) = p_{xx}(z)\hat{\mathbf{e}}_x\hat{\mathbf{e}}_x + p_{yy}(z)\hat{\mathbf{e}}_y\hat{\mathbf{e}}_y + p_{zz}(z)\hat{\mathbf{e}}_z\hat{\mathbf{e}}_z, \tag{20}$$

where its transverse components $p_{xx}(z) = p_{yy}(z)\ (= p_T(z))$ because of the symmetry. The normal component $p_N(z) = p_{zz}(z) = p_0$ must be a constant following the mechanical balance condition $\nabla \cdot \mathbf{p} = 0$. The constant, $p_0$, is the hydrodynamic pressure in either bulk phase. For a low-viscosity fluid, we can ignore the last term in Eq. (19), since $\tau - \frac{1}{2} \sim \nu \ll 1$. The normal component of the pressure tensor, $p_{zz}$, can then be further reduced to

$$p_{zz} = p_0 = \frac{1-d_0}{D}c^2 n + \frac{c^2 b\mathcal{G}}{2D}\psi^2 + \frac{3bc^4\mathcal{G}}{4D(D+2)}\psi\left[\psi''\left(\frac{dn}{dz}\right)^2 + \psi'\frac{d^2n}{dz^2}\right], \tag{21}$$

where $\psi' = d\psi/dn$ and $\psi'' = d^2\psi/dn^2$. In either of the liquid and gas phases far from the interface, the pressure $p_0$ satisfies the following relation

$$p_0 = \frac{1-d_0}{D}c^2 n_l + \frac{c^2 b\mathcal{G}}{2D}\psi^2(n_l) = \frac{1-d_0}{D}c^2 n_g + \frac{c^2 b\mathcal{G}}{2D}\psi^2(n_g). \tag{22}$$

This is simply the equation of state written for the two bulk phases. Eq. (21), coupled with Eq. (22) through the unknown constant $p_0$, is the equation from which the density profile $n(z)$ is to be solved with the boundary condition $dn/dz = 0$ at $z = \pm\infty$. A simple change of variable can reduce it further to obtain a formal solution. Let $(dn/dz)^2 = y$, and notice $\frac{d^2n}{dz^2} = \frac{1}{2}\frac{dy}{dn}$, then Eq. (21) will be transformed to the following first-order differential equation of $y$:

$$p_0 - \frac{1-d_0}{D}c^2 n - \frac{c^2 b\mathcal{G}}{2D}\psi^2 = \frac{3c^4 b\mathcal{G}}{8D(D+2)}\frac{\psi}{\psi'}\frac{d}{dn}(y\psi'^2), \tag{23}$$

where $\psi(n)$ is a known function of $n$ and all the derivatives are with respect to $n$. By direct integration, we obtain the following solution:

$$y(n) = \frac{8D(D+2)}{3c^4 b\mathcal{G}\psi'^2} \int \left( p_0 - \frac{1-d_0}{D} c^2 n - \frac{c^2 b\mathcal{G}}{2D} \psi^2 \right) \frac{d\psi}{\psi} \tag{24}$$
$$= \frac{4(D+2)}{3c^2\psi'^2} \left[ \left( \frac{2Dp_0}{c^2 b\mathcal{G}} - \frac{2(1-d_0)}{b\mathcal{G}} n \right) \ln\psi - \frac{\psi^2}{2} + \frac{2(1-d_0)}{b\mathcal{G}} \int \ln\psi\, dn \right],$$

where the pressure $p_0$ and the integration constant will be determined by Eq. (22) and the boundary conditions are $y(n_l) = y(n_g) = 0$. To be compatible with these boundary conditions, we must have

$$\int_{n_g}^{n_l} \left( p_0 - \frac{1-d_0}{D} c^2 n - \frac{c^2 b\mathcal{G}}{2D} \psi^2 \right) \frac{\psi'}{\psi} dn = 0. \tag{25}$$

From Eqs. (25) and (22), we are able to determine the pressure $p_0$ and the densities $n_l$ and $n_g$ in the bulk liquid and gas phases for any given parameter $(1-d_0)/\mathcal{G}$. This defines the coexistence curve of this LBE model.

Once the pressure, $p_0$, and the densities, $n_g$ and $n_l$, are determined, $y(n) = (dn/dz)^2$ is completely known. The function $y(n)$, and the densities $n_g$ and $n_l$, depend only upon a single parameter $(1-d_0)/\mathcal{G}$. The density profile $n(z)$ across the liquid-gas interface is then given by the following equation:

$$\int_{(n_g+n_l)/2}^{n} y(n)^{-1/2} dn = z, \tag{26}$$

which in general can not be expressed in terms of elementary functions except for some very special choices of $\psi(n)$.

The calculation of the surface tension is straight forward. By definition [18], in the case of a flat interface, the surface tension can be calculated from the components of the pressure tensor as the following integral across the interface

$$\sigma = \int_{-\infty}^{\infty} (p_0 - p_T) dz. \tag{27}$$

Using the expressions for the pressure tensor given by Eq. (13), we have

$$\sigma = \frac{c^4 b\mathcal{G}}{2D(D+2)} \int_{-\infty}^{\infty} \psi \frac{d^2\psi}{dz^2} dz. \tag{28}$$

Integrating by parts and considering the boundary condition that $d\psi/dz = 0$ at $z = \pm\infty$, we obtain the following equation

$$\frac{\sigma}{\mathcal{G}} = -\frac{c^4 b}{2D(D+2)} \int_{n_g}^{n_l} \psi'^2 [y(n)]^{1/2} dn, \tag{29}$$

where the right hand side is a function of $(1 - d_0)/\mathcal{G}$ only. Since $n(z)$ also depends only on $(1 - d_0)/\mathcal{G}$, we are able to change the surface tension independently of the density profile.

It is interesting to compare the above results with the classic thermodynamic theory of phase transition in PVT systems [19,20]. If we take Eq. (11) as an isotherm of a PVT system, the Maxwell equal-area construction (which requires $\int_{v_l}^{v_g} (p_0 - p) dv = 0$, where $v \sim 1/n$ is the molar volume and $v_l$ and $v_g$ are its values in the liquid and gas phases) yields the following equation for the coexistence curve,

$$\int_{n_g}^{n_l} \left( p_0 - \frac{1 - d_0}{D} c^2 n - \frac{c^2 b \mathcal{G}}{2D} \psi^2 \right) \frac{1}{n^2} dn = 0. \tag{30}$$

Compared with Eq. (25), we find that the coexistence curve of the LBE model will not agree with the thermodynamic theory unless we chose the function $\psi$ to have the form

$$\psi(n) = \psi_0 \exp(-n_0/n), \tag{31}$$

where $\psi_0$ and $n_0$ are arbitrary constants. This is not surprising since there is not an energy-conservation relation guaranteed by the kinetic equation and a well-defined temperature in this LBE model. Therefore the LBE model does not necessarily correspond to an exact isothermal process. Nevertheless, if we indeed choose $\psi$ to have the form of Eq. (31), the behavior of the LBE model will be consistent with that of an isothermal process, and the parameter $(1 - d_0)/\mathcal{G}$ can be used as a temperature scale. We can also derive from the "isotherm" (11) an effective Gibbs potential which is always minimum near phase transition, even though we do not have energy conservation in the model.

## IV. SIMULATION VERIFICATION

In this section, we verify the results of Sec. III by solving Eqs. (22), (24), (25), (29) numerically and comparing the solutions with simulation results. We chose the function $\psi(n)$ to be $1 - \exp(-n)$, the same as in the previous paper [16]. On a two-dimensional hexagonal lattice, the critical point of the system can then be calculated from Eq. (12) as $-(1-d_0)/\mathcal{G} = 1.5$, at the critical density $n = \ln 2 \simeq 0.6931$.

Numerically solving the density profiles from the equations derived in last section is a complicated procedure. Firstly for a given value of $-(1-d_0)/\mathcal{G}$, the integral on the left hand side of Eq. (25) is a function of $p_0$, where the integration limits $n_g$ and $n_l$ have to be solved from Eq. (22) for given $p_0$. We can then solve $p_0$ together with $n_g$ and $n_l$ by finding the root of this function. The NAG routine, C05AGE, was utilized to find the root of an arbitrarily given function. Once $n_g$, $n_l$ and $p_0$ are obtained, $y(n)$ was found by integrating Eq. (24) using the IMSL routine, QDAGS. The density profile, $n(z)$, is then easily found using $(dn/dz)^2 = y$ and employing IMSL routine, IVPBS. Integrating Eq. (29), we obtain the surface tension. The whole calculation takes a few seconds on a Cray-YMP C90.

The simulations were carried out on a two-dimensional hexagonal lattice. In most of the calculations, the collision time, $\tau$, is chosen to be 0.6 unless otherwise specified. We first verify the numerically calculated coexistence curve and the density profiles by simulations on a $64 \times 256$ lattice. Periodic boundary conditions are used in both directions. The initial density distribution was so set up that the density in half of the domain is higher than that in the other half. When the system reaches equilibrium after about $10^6$ iterations, a flat interface parallel to the shorter edge of the rectangular domain forms in the middle. The densities in the bulk phases are then measured for different values of $-(1-d_0)/\mathcal{G}$ and plotted in Fig. 1. The numerical solution of Eqs. (25), shown as the solid line, is found to be in very good agreement with simulation results. In Fig. 1, we also demonstrate the deviation from the coexistence curve predicted by the Maxwell construction, shown as the dashed line. This curve is obtained by solving Eq. (30) instead of Eq. (25) with the same procedure. The difference is small but noticeable, especially when $-(1-d_0)/\mathcal{G}$ goes far below the critical value. Of cause, this difference will not exist if the effective mass $\psi(n)$ is chosen to have the form given in Eq. (31).

The density profiles for several different values of the temperature-like parameter below its critical value are displayed in Fig. 2. To examine the isotropy of the surface tension, for each value of $-(1-d_0)/\mathcal{G}$, we performed two simulations in which the angles between the normal vector of the interface and the lattice links are 30° and 0° respectively [15]. The measured density profiles in both cases are displayed. No significant difference between these two cases is found. The isotropy of surface tension in this model is also evident as the previously displayed bubbles [16] are circular. Here, all the measurements were made at one cross section and at one time without any kind of average. The lines shown in Fig. 2 are the density profiles solved from Eqs. (21) and (22). The analytical predictions are again in excellent agreement with the simulation results. It is also seen that when the temperature-like parameter $-(1-d_0)/\mathcal{G}$ approaches the critical point, the distinctions between the two phases become smaller and will eventually vanish at the critical temperature.

Plotted in Fig. 3 is the surface tension scaled by the factor $1/\mathcal{G}$, numerically evaluated using Eq. (29), as a function of $-(1-d_0)/\mathcal{G}$. It is a monotonically decreasing function which goes to zero at the critical point. The agreement with the Laplace law was verified by simulation on a $128\times128$ lattice. For initial conditions, we set up a circular high-density region in the center of the domain. When the system reaches equilibrium, a liquid bubble in very good circular shape forms and the bubble radius and the pressure difference inside and outside the bubble are then measured. According to the Laplace law, for a 2-D circular bubble, the pressure difference $p_{in}-p_{out}$ and the surface tension $\sigma$ are related by

$$p_{in}-p_{out}=\frac{\sigma}{R}, \tag{32}$$

where $R$ is the radius of the bubble. Plotted in Fig. 4 are the measurements and straight lines drawn with the slopes given by the numerical solution of Eq. (29). The two cases are for $-(1-d_0)/\mathcal{G}=1.3$ and 1.4 respectively. The agreement between the simulation and the analytical results is good.

As a simple application, we compute the dispersion relation of the capillary waves. In the absence of gravity, at long wavelength limit, the dispersion relation of capillary wave on a free surface is (see, e.g., Ref. [22]):

$$\omega^2=\left(\frac{\sigma}{\rho}\right)k^3, \tag{33}$$

where $\sigma$ is the surface tension and $\rho$ is the density of the fluid. The dispersion relation given by the lattice Boltzmann model was measured by examining a standing wave at a liquid-gas interface.

The simulations were carried out on a rectangular domain, measuring $L \times H$ in lattice unit, with periodic boundary conditions in both directions. Initially a interface parallel to the shorter edge was set up in the middle of the domain. A single sinusoidal wave with wavelength $L$ was imposed on the interface and the amplitude of it was subsequently measured. To make the effects of finite water depth negligible, the aspect ratio of the domain, $H/L$ has to be large and was chosen to be $4 : \frac{\sqrt{3}}{2}$ here. The maximum value of the wave number $k = 2\pi/L$ that can be simulated is limited by the fact that the wave amplitude has to be much larger than the thickness of the interface but much smaller than $L$. Four simulations with wavelength $L = 256\sqrt{3}$, $128\sqrt{3}$, $64\sqrt{3}$ and $32\sqrt{3}$ were performed. The angular frequencies, $\omega$, are plotted in Fig. 5 versus wave numbers $k$, together with Eq. (33). From the plot, the comparison is quite satisfactory. In these simulations, the parameters are $-(1 - d_0)/\mathcal{G} = 1.1$, $d_0 = 0.5$, and $\tau = 0.6$. The surface tension and the densities of the two phases are $\sigma = 0.40$, $n_g = 0.063$, and $n_l = 2.23$.

## V. DISCUSSION

We have described in detail a lattice Boltzmann model for simulating fluids obeying a non-ideal gas equation of state. The equation of state can be changed arbitrarily. With a properly chosen equation of state, the LBE model can undergo a liquid-gas phase transition. In addition to the equation of state, we gave the bulk densities and the density profile at equilibrium, all as functions of a single temperature-like parameter. The surface tension is also given by numerical integration and it can be varied independent of the above equilibrium properties.

This LBE model provides an efficient method for simulating flows involving interfaces and phase transitions. Phenomena near a liquid-gas critical point can also be simulated. The advantage over other LGA or LBE models developed for similar purpose are apparent, since in addition to its computational efficiency, almost everything with this model is known and can be changed to fit any desired properties.

The collision time, $\tau$, should be made to depend on the local density when simulating a fluid system with large density variation, so that the two phases will have different transport coefficients.

The equilibrium properties discussed in this paper will not be altered by this change.

As we have pointed out before, the long-range interactions in systems with multiple components can also be treated in LBE model using the same scheme. The phase diagrams of such a system are more complicated and remain to be calculated. In general, work similar to that contained in this paper can be carried out for the multi-component system as well. This is necessary before this method can be applied to practical problems to produce accurate quantitative results.

The major weak point of this model, we believe, is the lack of an energy conservation relation and a dynamic temperature equation, in spite of the fact that a static temperature can be identified with a particular choice of the effective mass. Although lattice Boltzmann models with more speeds can be constructed to enable energy conservation [21], it is difficult to incorporate into it, at least with the scheme discussed here, the correct non-local inter-particle forces and conserve the total energy at the same time. This is mostly due to discrete lattice effect, that each time step is separated into a "streaming" and a "collision" step. It is difficult to keep the total energy conserved in these separated steps.

## ACKNOWLEDGEMENTS

We thank Dr. Gary Doolen and Dr. David Montgomery for helpful discussion and suggestions. This work was supported is part at Dartmouth by subcontract 9-XA3-1416E from Los Alamos National Laboratory and by NMT-5 at Los Alamos. The computation was performed in part using the resources located at the Advanced Computing Laboratory of Los Alamos National Laboratory, Los Alamos, NM 87545, and at the National Energy Research Supercomputer Center.

## FIGURES

FIG. 1. Coexistence curve when the function $\psi$ is chosen to be $1 - \exp(-n)$. The critical point is at $-(1 - d_0)/\mathcal{G} = 1.5$ and $n_c = \ln 2 \simeq 0.693$. The solid line is the solution of Eq. (25) and the dashed line is the solution to Eq. (30), the Maxwell construction. Both of them are obtained numerically. The diamonds are the results of numerical simulations.

FIG. 2. The density profiles across the liquid-gas interface. Solid lines are solved from Eq. (24) by numerical integration. For each value of $-(1 - d_0)/\mathcal{G}$, results of two simulations with the angle between the normal vector of the interface and the lattice links being 30° and 0° are shown. No significant difference is found between them. As $-(1 - d_0)/\mathcal{G}$ approaches the critical value, the distinction between the two phases becomes negligible.

FIG. 3. Surface tension, scaled by $1/\mathcal{G}$, as function of $-(1-d_0)/\mathcal{G}$. This curve is obtained by numerical integration using Eq. (29).

FIG. 4. Laplace law tested by numerical simulations. The marks indicate measured pressure differences inside and outside a 2-D circular bubble $p_{in} - p_{out}$, versus the reciprocal of the measured bubble radii. The lines are the analytic predictions. The two cases are for $-(1 - d_0)/\mathcal{G} = 1.3$ and $-(1 - d_0)/\mathcal{G} = 1.4$.

FIG. 5. Dispersion relation of capillary wave at a liquid-gas interface. The diamonds are measured from lattice Boltzmann simulations and the dashed straight line is given by Eq. (33).

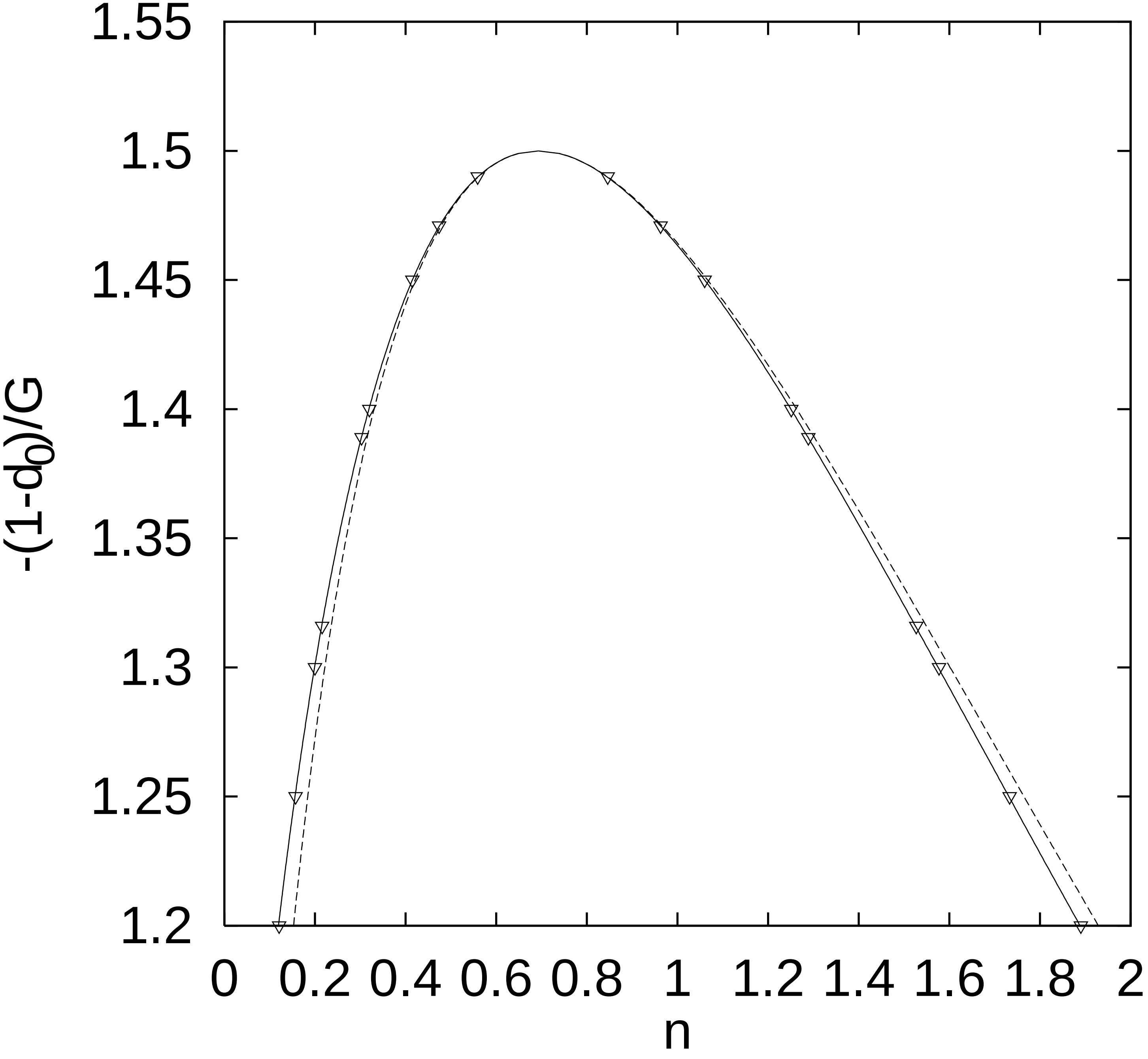

Figure 1

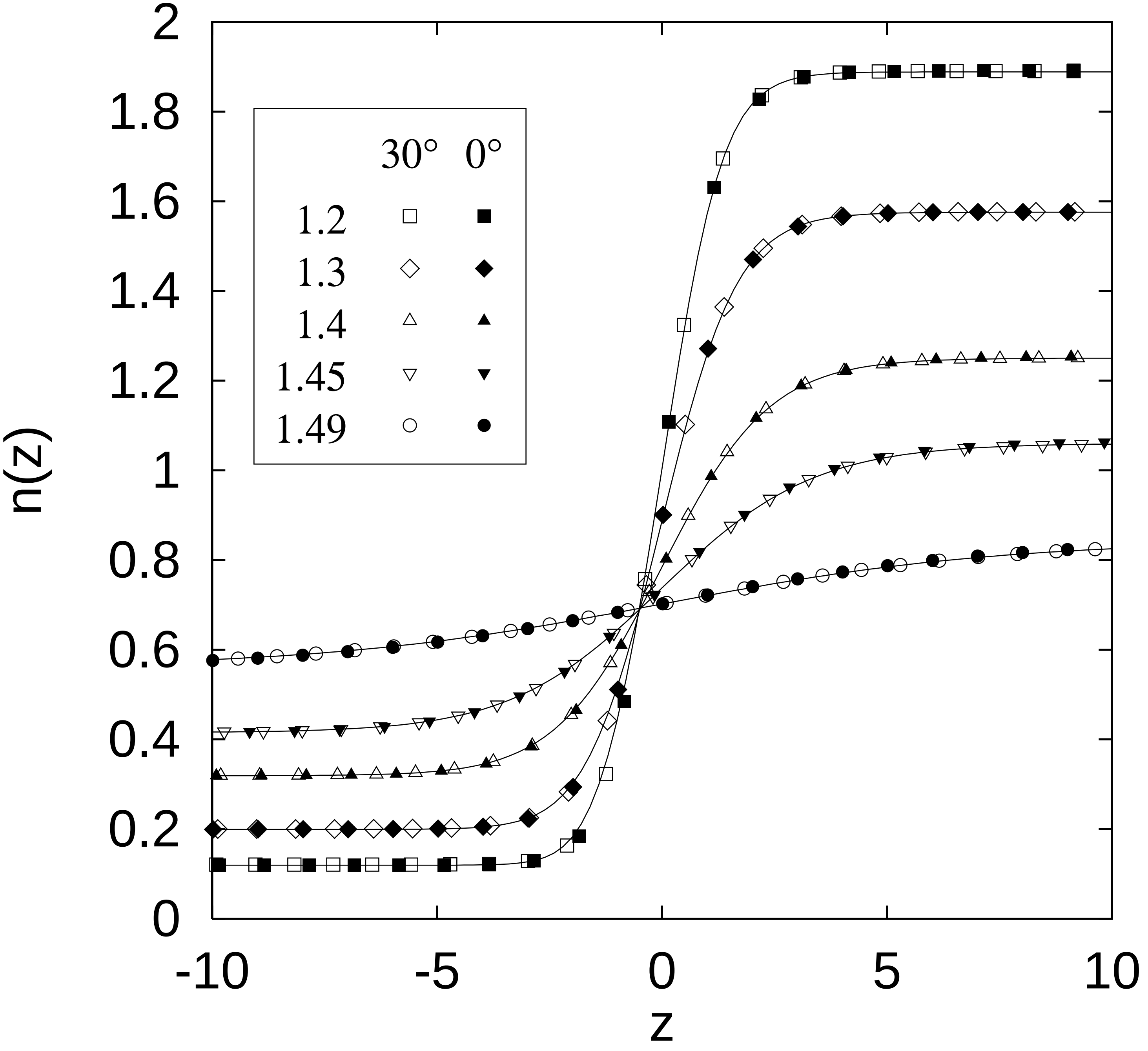

Figure 2

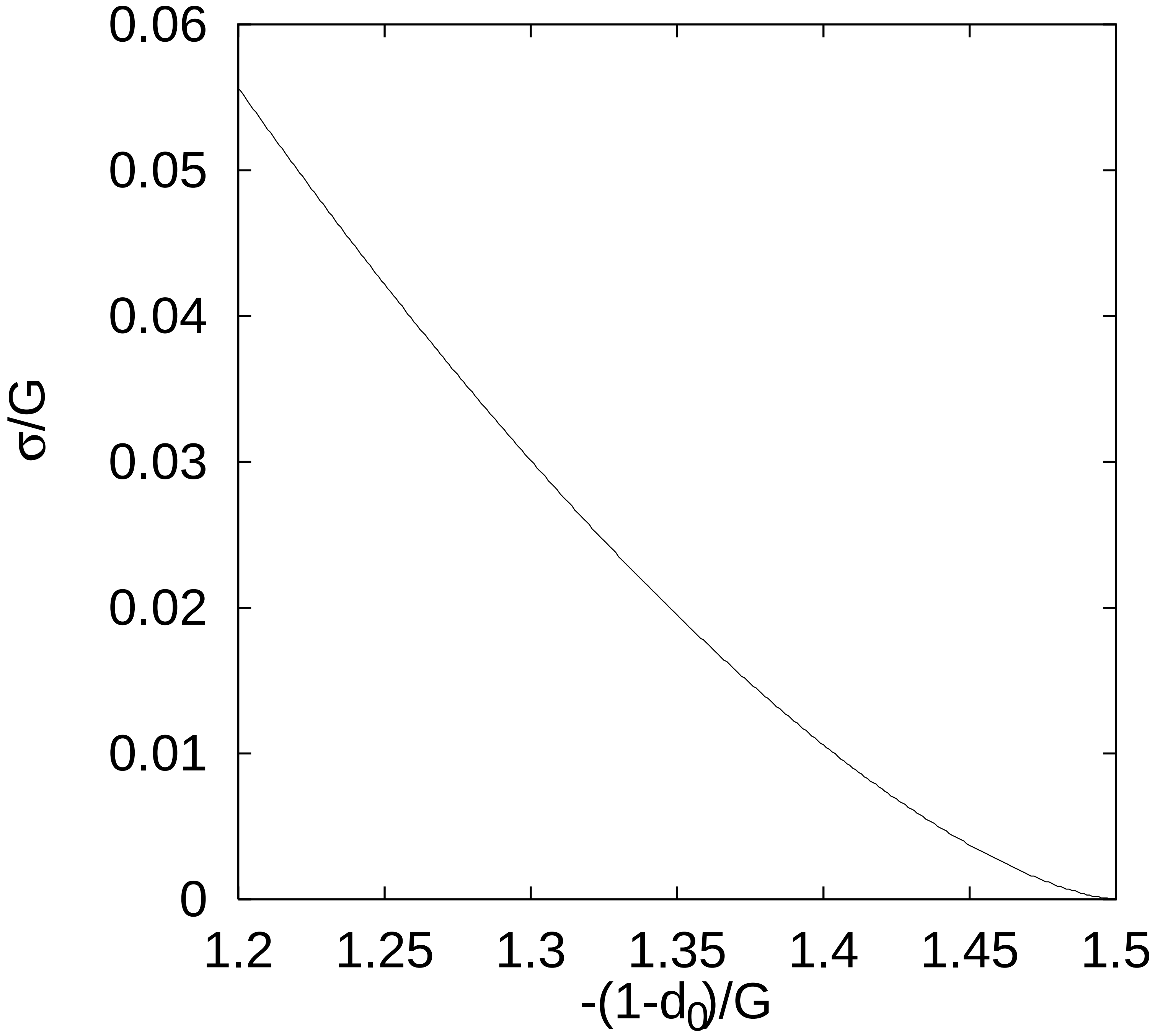

Figure 3

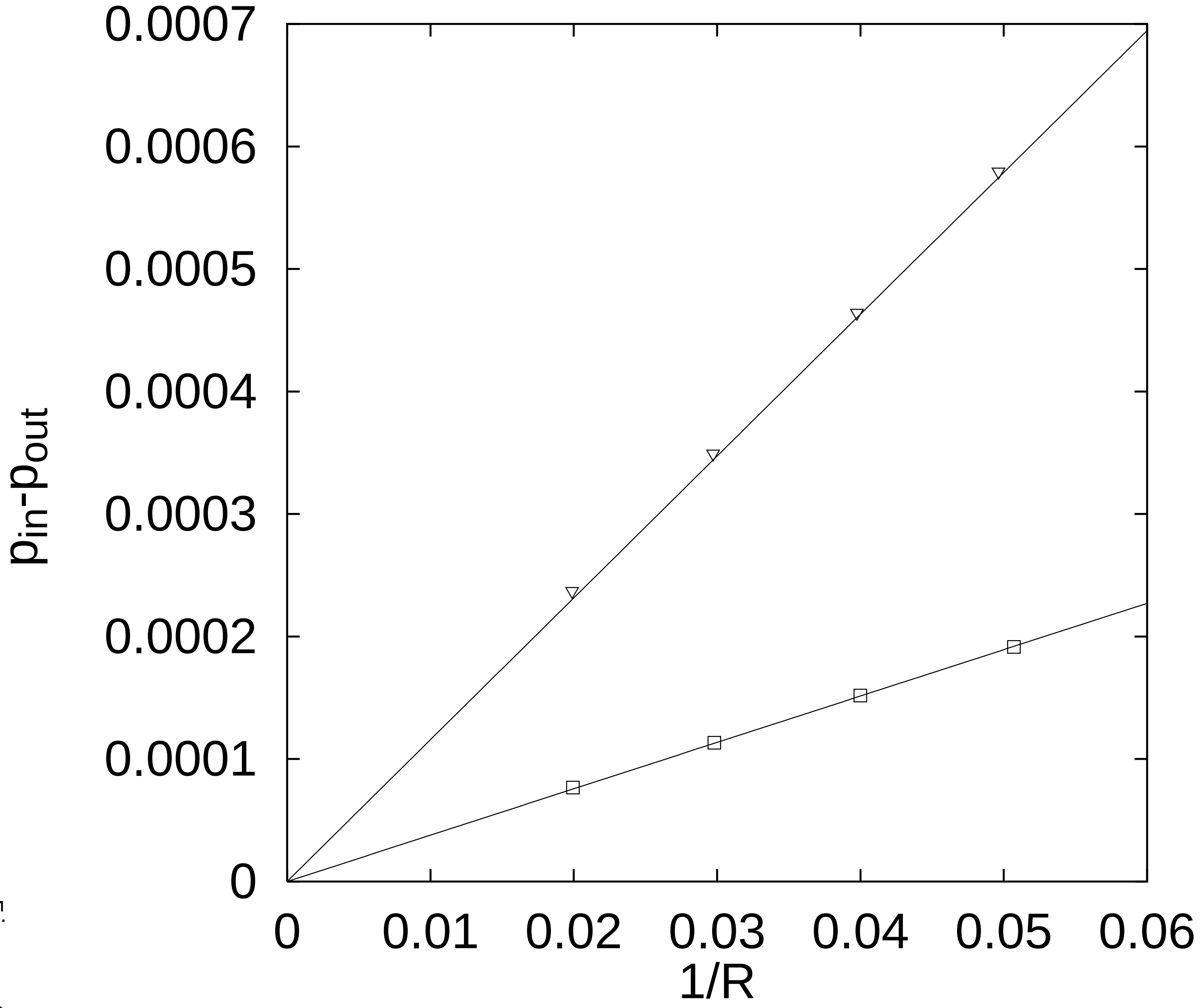

Figure 4

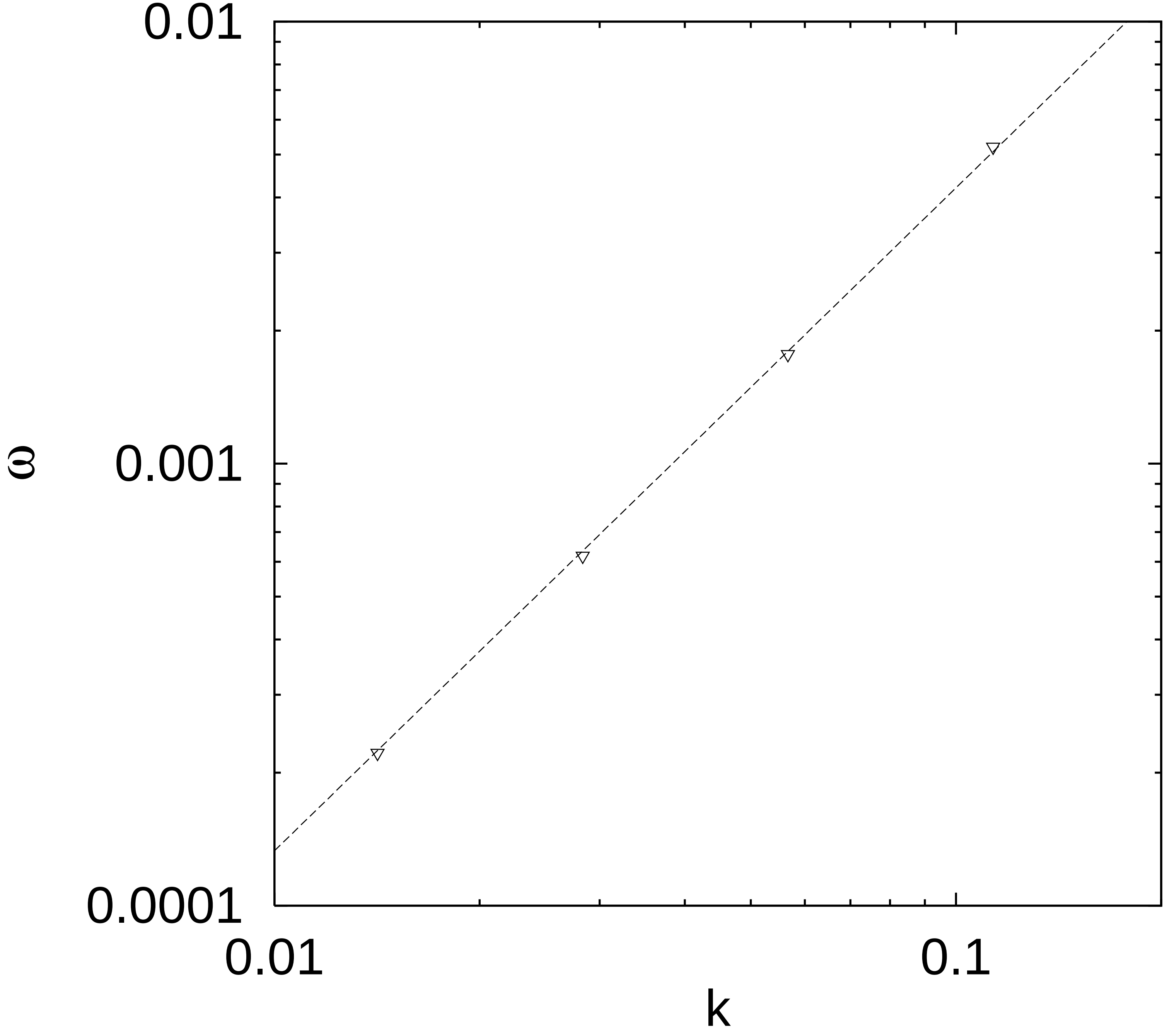

Figure 5